\documentclass{article}
\usepackage{setspace}
\usepackage{arxiv}

\usepackage[utf8]{inputenc} 
\usepackage[T1]{fontenc}    
\usepackage{hyperref}       
\usepackage{url}            
\usepackage{booktabs}       
\usepackage{amsfonts}       
\usepackage{nicefrac}       
\usepackage{microtype}      
\usepackage{float}          
\usepackage{graphicx}
\usepackage{cite} 
\usepackage{doi}
\usepackage{subfiles}
\usepackage{amsmath}
\usepackage{bm} 
\usepackage{subcaption}
\usepackage{xcolor}
\usepackage{ulem}

\usepackage[mathlines]{lineno}

\title{Deep learning forecasts the spatiotemporal evolution \\of fluid-induced microearthquakes}

\author{%
  \href{https://orcid.org/0000-0003-2960-4601}{\includegraphics[scale=0.06]{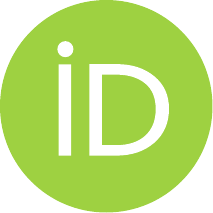}\hspace{1mm}Jaehong Chung}\textsuperscript{1} \quad
  \href{https://orcid.org/0000-0003-3286-4682}
  {\includegraphics[scale=0.06]{orcid.pdf}\hspace{1mm}Michael Manga}\textsuperscript{2} \quad
  \href{https://orcid.org/0000-0002-3926-8587}
  {\includegraphics[scale=0.06]{orcid.pdf}\hspace{1mm}Timothy Kneafsey}\textsuperscript{3} \quad
  \href{https://orcid.org/0000-0003-1711-1850}{\includegraphics[scale=0.06]{orcid.pdf}\hspace{1mm}Tapan Mukerji}\textsuperscript{1,4}
  \href{https://orcid.org/0000-0002-8853-2022}
  {\includegraphics[scale=0.06]{orcid.pdf}\hspace{1mm}Mengsu Hu}\thanks{Corresponding author: mengsuhu@lbl.gov}\textsuperscript{3, *} \quad  
  \\[8pt] 
  \textsuperscript{1}Department of Geophysics, Stanford University, Stanford, CA, USA  \\[3pt]  
  \textsuperscript{2}Department of Earth and Planetary Science, University of California Berkeley, Berkeley, CA, USA   \\[3pt]  
  \textsuperscript{3}Energy Geoscience Division, Lawrence Berkeley National Laboratory, Berkeley, CA, USA   \\[3pt]    
  \textsuperscript{4}Department of Energy Science and Engineering, Stanford University, Stanford, CA, USA  
}

\date{}

\renewcommand{\undertitle}{ }
\renewcommand{\shorttitle}{ } 

\hypersetup{
pdftitle={A template for the arxiv style},
pdfsubject={q-bio.NC, q-bio.QM},
pdfauthor={David S.~Hippocampus, Elias D.~Striatum},
pdfkeywords={First keyword, Second keyword, More},
}

\begin{document}
\maketitle

\begin{abstract}
Microearthquakes (MEQs) generated by subsurface fluid injection record the evolving stress state and permeability of reservoirs. Forecasting their full spatiotemporal evolution is therefore critical for applications such as enhanced geothermal systems (EGS), CO$_2$ sequestration and other geo-engineering applications. We present a transformer-based deep learning model that ingests hydraulic stimulation history and prior MEQ observations to forecast four key quantities: cumulative MEQ count, cumulative logarithmic seismic moment, and the 50th- and 95th-percentile extents ($P_{50}, P_{95}$) of the MEQ cloud. Applied to the EGS Collab Experiment 1 dataset, the model achieves $R^2 >0.98$ for the 1-second forecast horizon and $R^2 >0.88$ for the 15-second forecast horizon across all targets, and supplies uncertainty estimates through a learned standard deviation term. These accurate, uncertainty-quantified forecasts enable real-time inference of fracture propagation and permeability evolution, demonstrating the strong potential of deep-learning approaches to improve seismic-risk assessment and guide mitigation strategies in future fluid-injection operations.
\end{abstract}


\section{Introduction}\label{sec:intro}
Subsurface applications for climate mitigation and sustainability are essential to achieving the net-zero emissions target set by the Intergovernmental Panel on Climate Change (IPCC) for 2050 \cite{metz2005ipcc}. Key geo-engineering strategies include the development of enhanced geothermal systems (EGS) for renewable energy generation and the geological storage of carbon dioxide (CO$_2$) to reduce atmospheric greenhouse gas concentrations. The U.S. Geological Survey (USGS) estimates that EGS could provide over 500 GWe of electricity in the western United States alone \cite{williams2008assessment}. In addition, carbon dioxide sequestration has the potential to store at least 1,000 GtCO$_2$ in saline aquifers, with further storage capacity available in depleted oil and gas reservoirs and coal formations \cite{bachu2003sequestration, damen2006health}. Despite the immense potential to reduce greenhouse gases through these subsurface applications, a key challenge is the induced seismicity that can result from fluid injection operations \cite{rutqvist2012geomechanics, yeo2020causal, ellsworth2019triggering}. Fluid injection perturbs in-situ stress fields in the subsurface, potentially leading to the reactivation of pre-existing faults or the creation of new fractures, both potentially compromising the integrity of reservoirs \cite{wang2017induced}. Notable examples include the magnitude 5.7 2015 Prague and magnitude 5.8 2016 Pawnee earthquakes in Oklahoma after  wastewater injection \cite{rajesh2021characterization, johann2018surge, hincks2018oklahoma, manga2016increased}---and a magnitude 3.9 earthquake following circulation tests for the EGS project in Vendenheim, France \cite{fiori2023monitoring}. These events underscore the critical need for accurate forecasting of induced seismicity to ensure the safe implementation of subsurface technologies.

Accurately forecasting fluid-induced seismicity remains a significant challenge due to the complex interactions between geological, hydrological, and mechanical factors \cite{rutqvist2012geomechanics, shapiro2015fluid}. Traditional approaches rely on physics-based models to estimate induced seismicity by coupling fluid flow, mechanical deformation, and seismicity rates \cite{boyet2024forecasting, lu2021coupled, mcclure2011investigation, zhai2019pore}. Although these models can capture intricate subsurface interactions, they face significant limitations in real-world applications. Challenges include uncertainties in fracture geometries, material heterogeneity, and in-situ stress conditions. Moreover, assumptions such as isotropic material properties or idealized fracture networks are often required, reducing predictive accuracy. High computational costs associated with three-dimensional modeling with complex fracture geometries further restrict their use in practical forecasting and operational decision-making \cite{boyet2024forecasting, mcclure2011investigation}. As a result, discrepancies between modeled and observed seismicity frequently occur. 

From a statistical perspective, the Epidemic-Type Aftershock Sequence (ETAS) model provides a forecasting approach for both natural and fluid-induced seismicity, based on the assumption that an earthquake can trigger clusters of aftershocks \cite{kumazawa2014nonstationary,ritz2024pseudo}. In particular, nonstationary ETAS models have effectively demonstrated their capability in detecting the impacts of fluid-induced seismicity by employing a nonstationary background rate \cite{hainzl2005detecting, kumazawa2013quantitative, kumazawa2014nonstationary, petrillo2024fluids}. This capability positions ETAS as a valuable tool for generating probabilistic earthquake forecasts. 
However, determining key parameters, including the timing of peak activity, solely based on statistical analysis has been challenging \cite{aochi2021statistical}. Thus, successful applications of ETAS models to spatiotemporal forecasting of MEQs due to fluid injection may be limited.

Data-driven approaches---particularly machine learning---have emerged as powerful complements or alternatives to traditional frameworks including both physics-based and statistical approaches, in a range of geoscientific applications
\cite{qin2022forecasting,zhang2022application, chung2024prediction, camps2021deep, maniar2018machine, bergen2019machine, yu2021deep, mousavi2022deep, zhu2019phasenet, reichstein2019deep, anikiev2023machine, jinqiang2021review, mousavi2016seismic}. These methods do not require detailed prior knowledge of uncertain subsurface properties but instead leverage large datasets from monitoring systems to identify patterns and correlations that can be used for forecasting. For instance, Li et al. \cite{li2023physics} employed deep learning—with and without physical constraints—to forecast the seismicity rate, which was then used to estimate the maximum magnitude of fluid-induced microseismicity. Yu et al. \cite{yu2024crustal} utilized a bidirectional long short-term memory (Bi-LSTM) neural network to predict fluid-induced permeability evolution based on MEQ features, including seismic rate and cumulative logarithmic seismic moment. Li et al. \cite{li2022induced} employed an LSTM model to predict average permeability changes inferred from the seismicity data. Mital et al. \cite{mital2024modeling} used an LSTM model to predict pore pressure and associated fault displacements given the fluid injection cycles. These studies demonstrate that deep learning approaches can effectively capture the temporal evolution of permeability or micro-seismicity based on operational parameters. However, they often focus solely on temporal predictions without considering the spatial evolution of MEQs, which is critical for assessing the extent of affected areas and potential impacts. Furthermore, these models rely on simplified assumptions for permeability changes, such as the migration of the triggering front of the MEQ cloud assuming proportionality to the square root of time since the initiation of injection, which is inconsistent with observed MEQ data \cite{hummel2013nonlinear}. These idealizations limit the applicability and accuracy of the models in complex scenarios.

Our study advances the forecasting of the spatiotemporal evolution of MEQs induced by hydraulic stimulation using a deep learning approach that tackles these challenges. Specifically, we employ transformer networks, a type of neural network architecture that uses self-attention mechanisms to capture complex dependencies within data sequences \cite{vaswani2017attention, zeng2023transformers}. Compared with recurrent neural networks such as LSTMs, transformer networks can model long-range temporal dependencies more efficiently and are less susceptible to issues like vanishing gradients \cite{zhou2021informer}. Their ability to focus on different parts of the input data through attention mechanisms makes them particularly well-suited for capturing both spatial and temporal patterns in MEQ data. Based on hydraulic stimulation history, our model predicts key MEQ features, including the cumulative number of MEQs, cumulative seismic moment, and the spatial extent of induced micro-seismicity. By incorporating both spatial and temporal information, the model provides more comprehensive forecasts that can inform real-time monitoring and risk mitigation strategies in subsurface activities.

\section{Results}\label{sec:results}
\begin{figure}[h]
    \centering
    \includegraphics[width=1.0\textwidth]{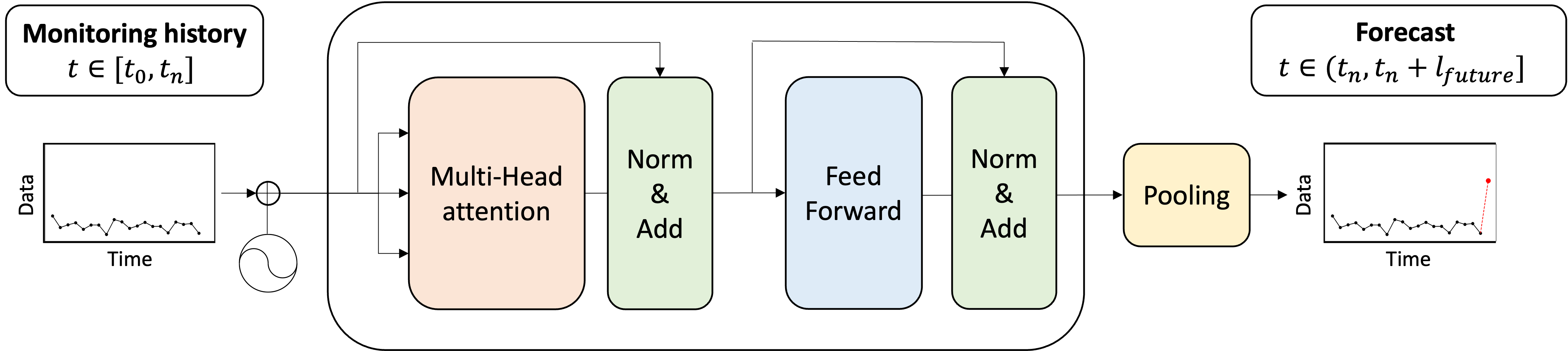}
    \caption{Architecture of the transformer-based MEQ forecasting model. Given input history from time steps $t_0$ through $t_n$, the model predicts MEQ features at future time steps $t_{n+1}$ through $t_{n} + l_{\mathrm{future}}$, where \(l_{\mathrm{future}}\) is the forecast range.}
    \label{fig:architecture}
\end{figure}

We use hydraulic stimulation data and MEQ history from the EGS Collab \cite{schoenball2020creation, fu2021close}. Figure~\ref{fig:architecture} shows the architecture of our transformer model for forecasting the spatiotemporal evolution of MEQs based on hydraulic stimulation and MEQ histories (see Section~\ref{sec:method}).

\subsection{EGS Collab hydraulic stimulation datasets }\label{sec:results_dataset}
We utilize hydraulic stimulation and MEQ data from the EGS Collab project, intermediate-scale (10-20 m) field tests at the Sanford Underground Research Facility (SURF) in Lead, South Dakota. This study focuses on the Experiment 1 data, aimed at producing a fracture network connecting an injection well to a production well via hydraulic fracturing \cite{kneafsey2018overview}. A series of stimulations and flow tests were conducted at a depth of 1.5 km to re-open and generate hydraulic fractures in crystalline rock under reservoir-like stress conditions, with passive seismic data cataloged \cite{qin2024source} and Continuous Active-Source Seismic Monitoring (CASSM) \cite{schoenball2020creation, feng2024monitoring, kneafsey2019egs}. 

\begin{figure}[h]
    \centering
    \includegraphics[width=1.0\textwidth]{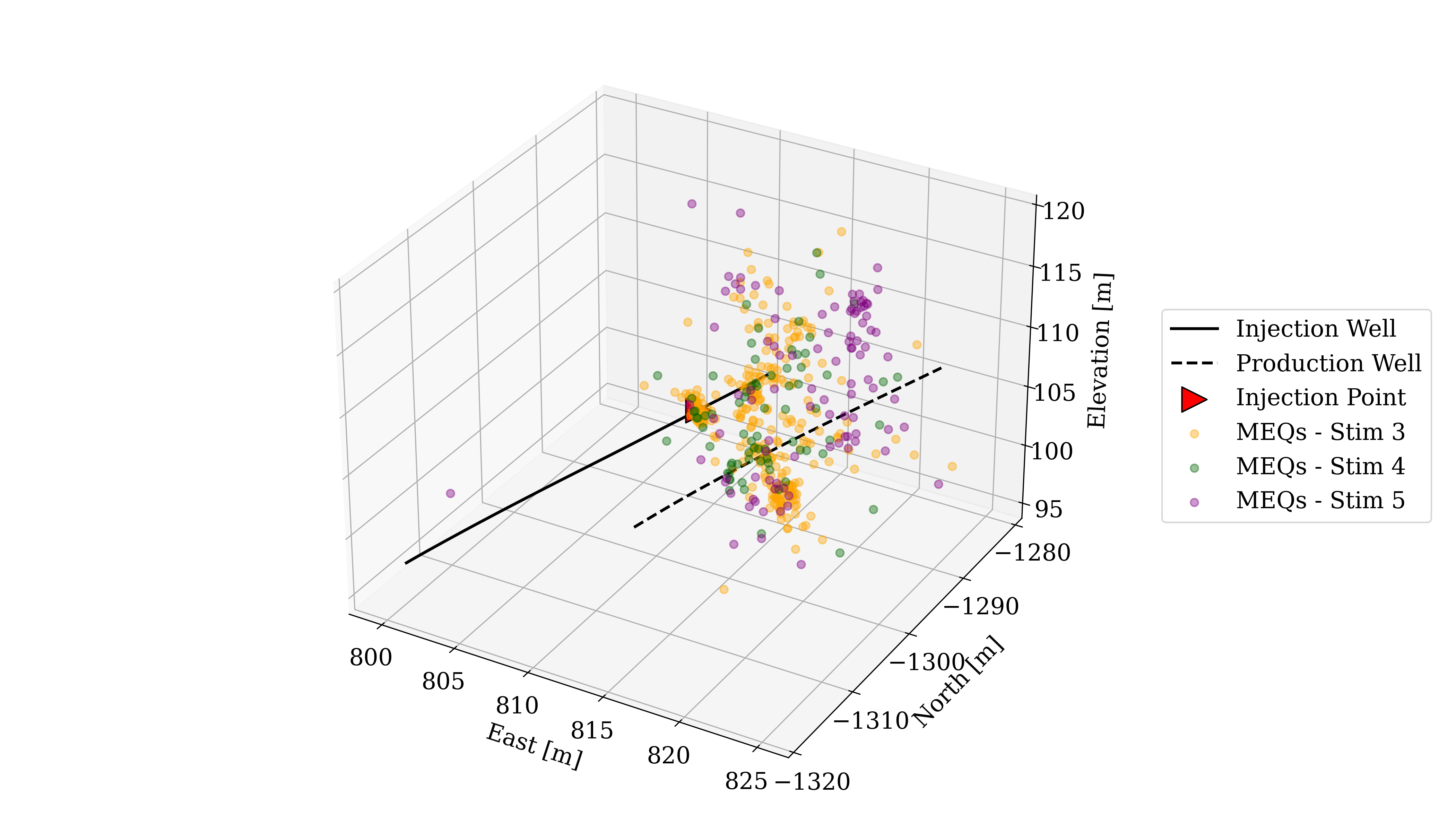}
    \caption{Fluid-induced microearthquakes for each hydraulic stimulation along with injection and production well. The red triangle denotes the injection point at the 50 m notch in the injection well. }\label{fig:data_stim3_5}
\end{figure}

Figure \ref{fig:data_stim3_5} shows the stimulation-induced MEQs for each stimulation event along with the injection and production wells. Two 60 m boreholes were used for injection (E1-I) and production (E1-P), respectively. A total of five stimulation episodes were carried out in May 2018. During the first two stimulations, injections at flow rates less than 1L/min produced few MEQs. In addition, water leakage was observed between the production well and one monitoring well. Thus, the injection point was moved to a notch at a depth of 50 m in the injection hole (red triangle in Figure \ref{fig:data_stim3_5}) starting from Stimulation 3 and used through Stimulation 5 . From Stimulations 3 to 5, three continuous hydraulic stimulations were performed using controlled step-rate injections to re-open or create fractures around the injection well, with the maximum injection rate reaching up to 5L/min, resulting in rich MEQ signals \cite{kneafsey2020egs, fu2021close}. Thus, this study uses data from Stimulations 3 to 5, generated from the same injection point with a rich MEQ history, to train neural networks. The data were recorded at 1-second intervals. Stimulations 3 and 4 each lasted approximately 1 hour (3600 time steps), and the first 1 hour and 10 minutes of Stimulation 5 were used (4100 time steps). These continuous records were segmented into overlapping input-output windows for supervised training, validation, and testing, as described in Section \ref{sec:dataPreprocessing}.

\begin{figure}[h]
    \centering
    \includegraphics[width=1.0\textwidth]{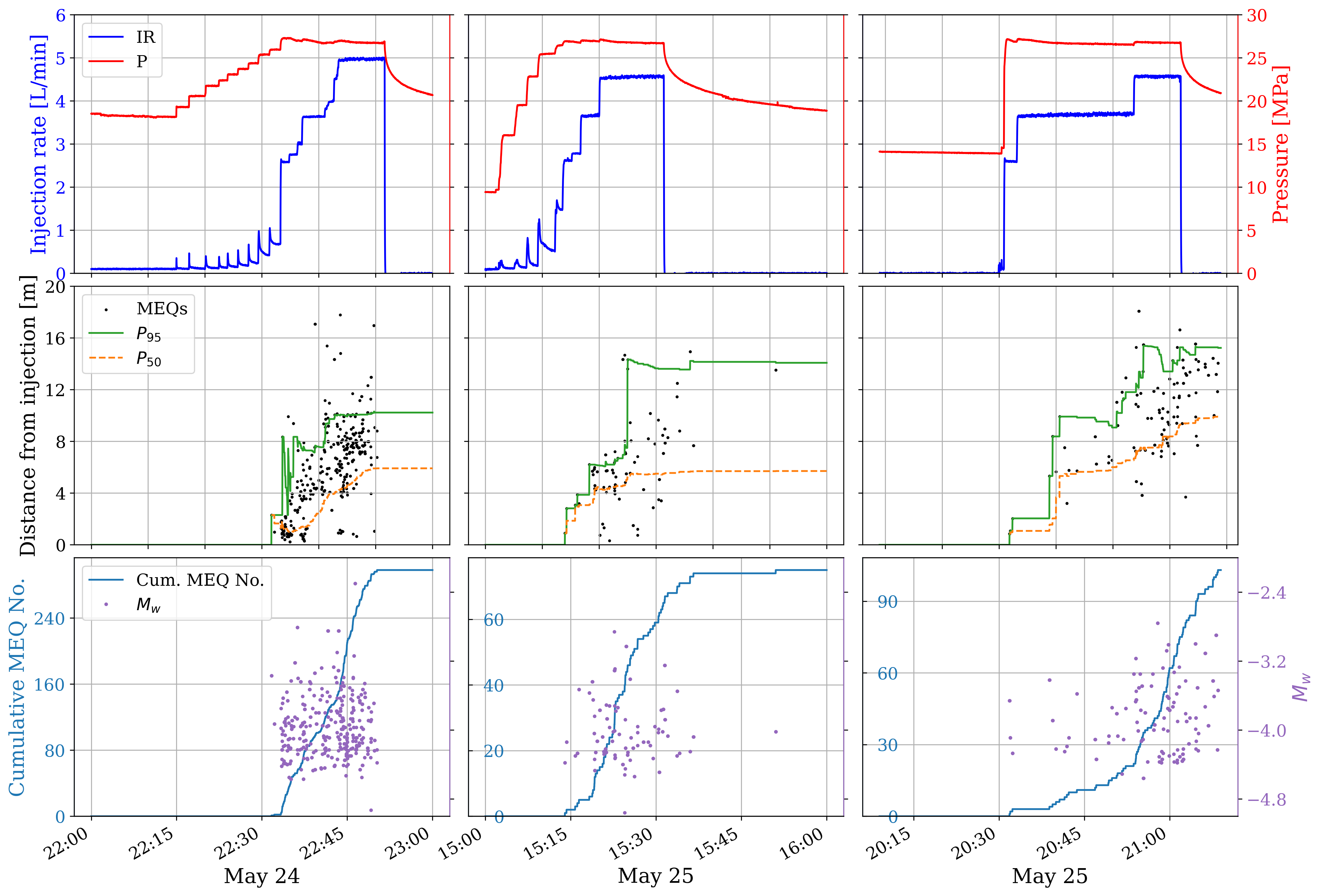}
    \caption{Microearthquake and injection history for EGS Collab Stimulation 3–5. Columns correspond to: Stimulation 3 (training data), Stimulation 4 (validation data), and Stimulation 5 (test data). The first row presents the hydraulic stimulation history, showing injection rate (blue) and injection pressure (red). The second row displays the locations of microearthquakes (MEQs) relative to the injection point, with distances calculated as the Euclidean distance between the injection point and observed microseismic events. $P_{95}$ and $P_{50}$ represent the 95th and 50th percentile distances over time. The third row shows the cumulative number of MEQ events and the magnitude of each discrete event.}
    \label{fig:stimulation_data}
\end{figure}

Figure \ref{fig:stimulation_data} presents the series of stimulations along with the spatiotemporal MEQ data and corresponding magnitudes. Detailed information about the MEQs—including location, time, and magnitudes—was continuously monitored during the hydraulic stimulations \cite{schoenball2020creation, fu2021close}. In addition, to quantify the spatial extent of MEQs in response to fluid injection, we extracted the 95th and 50th percentiles (median) distances of the MEQ clouds from the injection points as a function of time. Although the monitoring array is extensive, the catalog still carries intrinsic uncertainties: hypocenter locations are accurate to about 1 m and there is no reported uncertainty range for magnitude \cite{schoenball2020creation}. These uncertainties limit the fidelity of the training data and establish a floor on achievable forecast accuracy. Additionally, including all raw events---without excluding those below the magnitude of completeness---could constrain the neural network's capability to learn underlying MEQ patterns (Fig.~S1).

\subsection{Forecasting performance}\label{sec:results_performance}
\begin{figure}[h]
    \centering
    \includegraphics[width=0.8\textwidth]{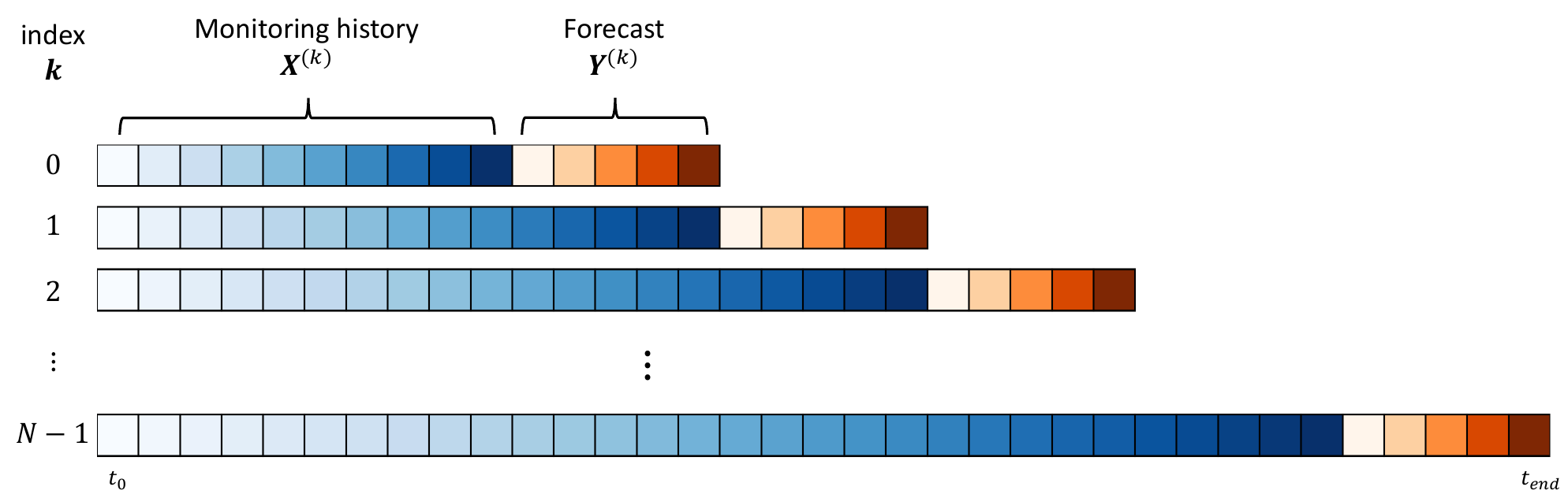}
    \caption{Schematic of the forecasting procedure. \(\bm{X}^{(k)}\) denotes the cumulative monitoring input and \(\bm{Y}^{(k)}\) the corresponding forecast window; \(k\) is the segment index. The forecasting range is $l_\text{future}$ (see Section \ref{sec:method})}
    \label{fig:forecastingSchematic}
\end{figure}

We evaluate three forecast intervals—1 s, 15 s, and 30 s—using a sliding-window strategy. At each forecasting instant $t_n$, the model ingests the entire monitoring history $[t_0,t_n]$ and predicts subsequent interval $[t_{n+1}, t_{n} + l_\text{future}]$, where $l_{\text{future}}$ is the forecast range (e.g., 1 s, 15 s, or 30 s). For instance, when using a 15 s range, the model forecasts the next 15 seconds (e.g., $t_{101}-t_{115}$) based on the history data $t_1 - t_{100}$. Once actual monitoring for these 15 seconds are recorded, these new data ($t_{101} - t_{115}$) are appended to the monitoring history. The model then uses the extended history $t_1 - t_{115}$ to forecast the following segment $t_{116} - t_{130}$, and this procedure repeats until the monitoring concludes. Since the model consistently utilizes actual measurements without recycling previously predicted outputs, forecasting errors do not accumulate over successive forecasts (Figure~\ref{fig:forecastingSchematic}).

\begin{figure}[h]
    \centering
    \includegraphics[width=1.0\textwidth]{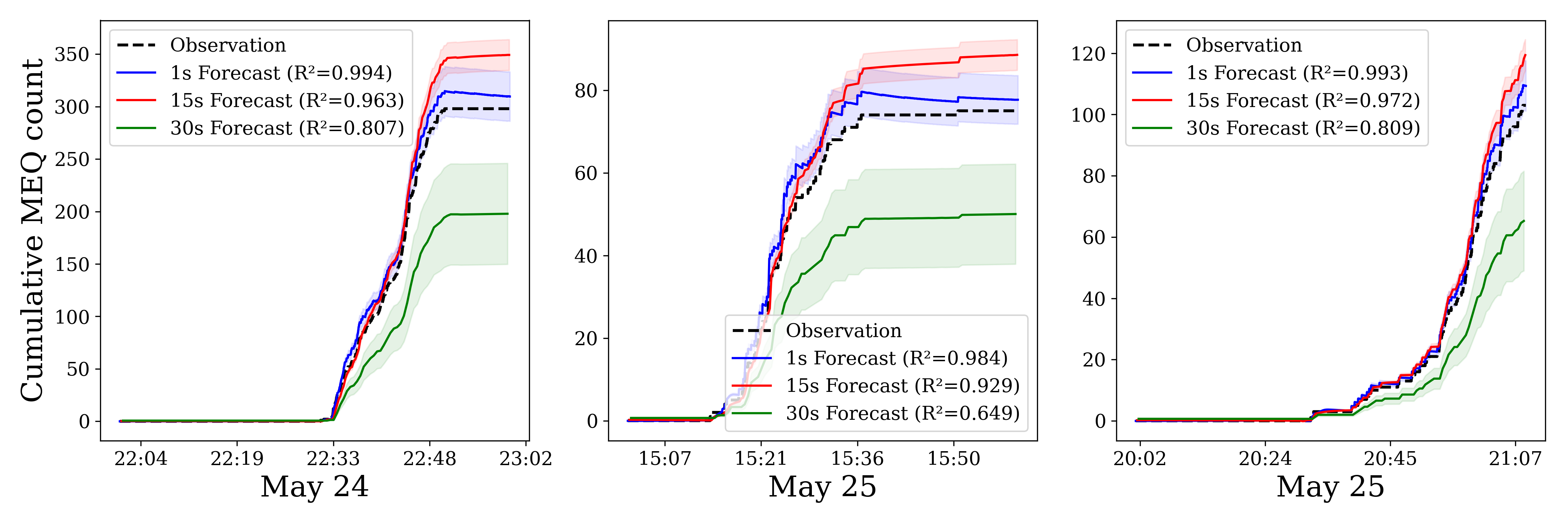}
    \caption{Cumulative MEQ counts: observed data (black dotted) versus forecasts for the 1-second (blue), 15-second (red), and 30-second (green) models. Each forecast curve is constructed by predicting successive, non-overlapping segments whose length equals the forecast interval and concatenating them to cover the full record. Panels show the training (left), validation (middle), and test (right) sets. Shaded bands denote \(\pm\sigma\) (one predicted standard deviation, corresponding to \(\approx68\%\) coverage under a Gaussian assumption).}
    \label{fig:Cum_MEQ_No}
\end{figure}

Figure~\ref{fig:Cum_MEQ_No} compares the forecasted and observed cumulative MEQ counts. For the 1-second forecast model the predicted curves are virtually indistinguishable from the ground truth, even on unseen data (validation $R^{2}=0.999$, test $R^{2}=0.980$). The 15-second forecast model maintains high fidelity (validation $R^{2}=0.929$, test $R^{2}=0.972$), with a slight tendency to overestimate MEQ growth during the most intense injection phases. The 30-second forecast model still captures the overall trend but systematically underpredicts the MEQ count late in each episode (validation $R^{2}=0.649$, test $R^{2}=0.809$). These results show that the transformer delivers excellent short-term forecasts, with accuracy declining gradually as the forecast window lengthens.

Second, we forecast the cumulative logarithmic seismic moment, a proxy for the activated reservoir volume and thus a key metric for planning new production wells \cite{rothert2010passive, baisch2009investigation}. Following Yu et al.~\cite{yu2024crustal}, the cumulative moment $\mathcal{M}$ is defined as
\begin{equation}
    \mathcal{M}(t_i)=\int_{t_0}^{t_i} \log M_0 \, dt,
\end{equation}
with
\begin{equation}
    \log M_0 = 1.5\,M_w + 13.5,
\end{equation}
where $M_0$ is the seismic moment, $M_w$ the moment magnitude, $t_0$ the start of injection, and $t_i$ the current injection time.

\begin{figure} 
    \centering
    \includegraphics[width=1.0\textwidth]{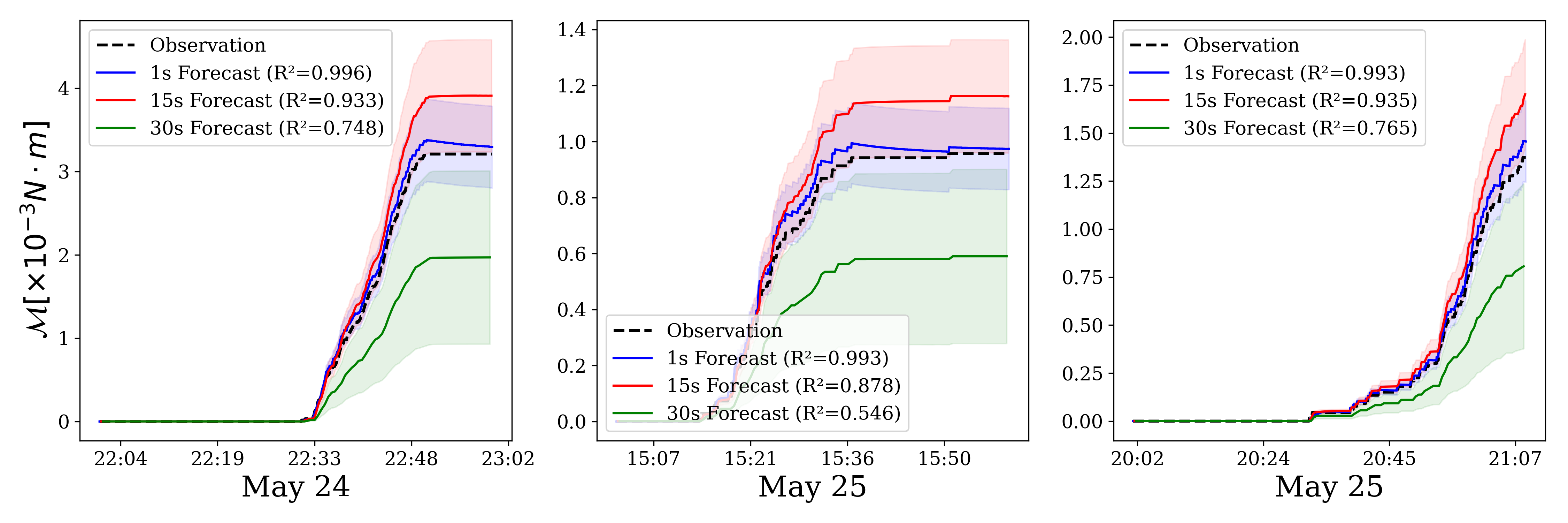}    
    \caption{Cumulative logarithmic seismic moment: observed data (black dotted) versus forecasts from the 1-, 15-, and 30-second models (blue, red, green) for the training (left), validation (middle), and test (right) sets. Shaded bands denote \(\pm\sigma\).}
    \label{fig:seismic_moment}
\end{figure}

Figure~\ref{fig:seismic_moment} compares the predicted and observed cumulative moments for the 1-, 15-, and 30-second forecast models across the three data splits. The 1-second forecast model reproduces the observations almost exactly (validation \(R^{2}=0.999\), test \(R^{2}=0.978\)). Performance remains high at 15-second forecast model (validation \(R^{2}=0.878\), test \(R^{2}=0.935\)), although the predictive bands widen compared with the 1-second case. At 30-second forecast the model still captures the overall trend but underestimates the released seismic energy (validation \(R^{2}=0.546\), test \(R^{2}=0.765\)). These results confirm that our neural network effectively links hydraulic-energy input to seismic-energy release, providing reliable short-term estimates of cumulative moment while showing a gradual and interpretable loss of accuracy as the forecast range increases.

\begin{figure}
    \centering
    \includegraphics[width=1.0\textwidth]{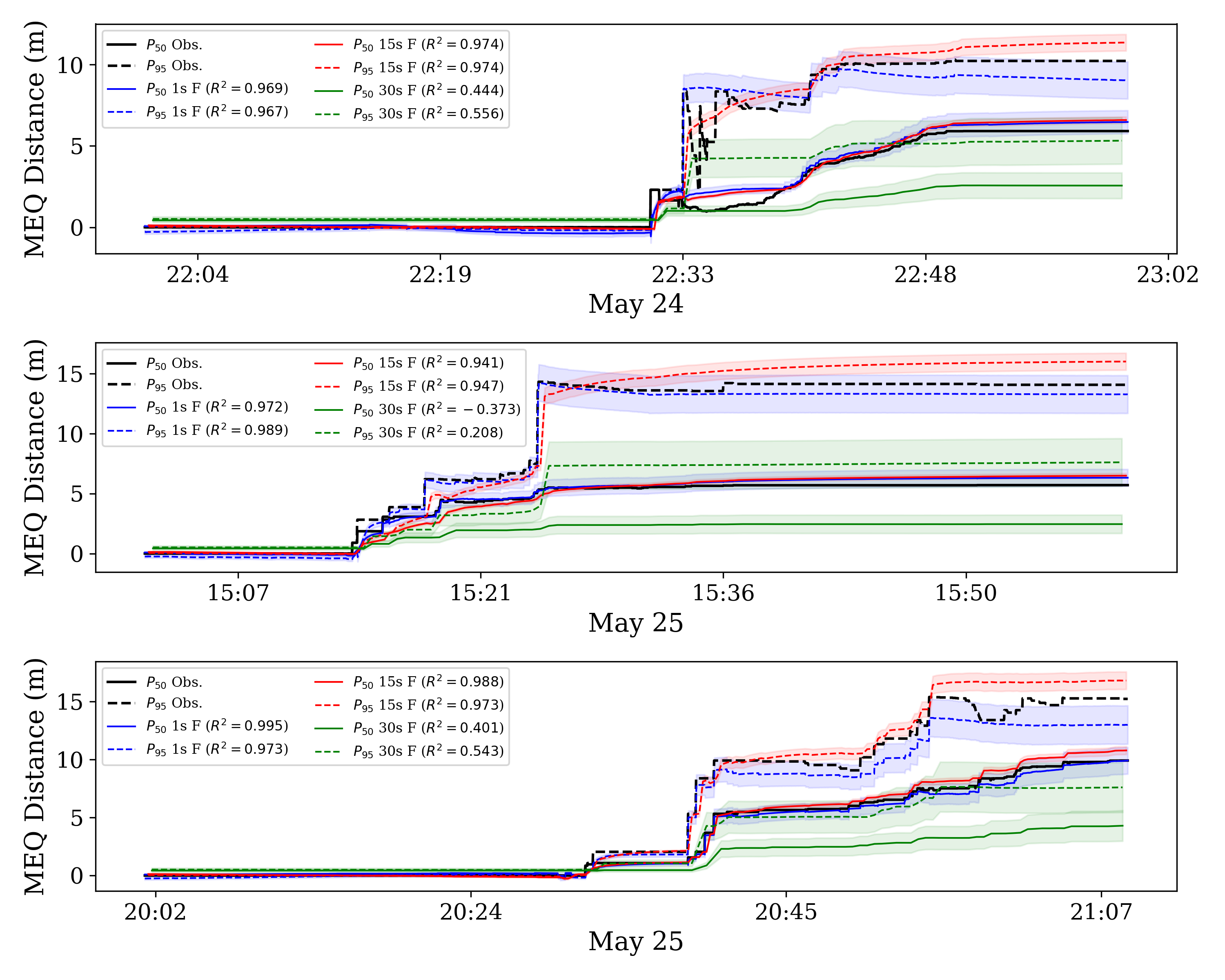}    
    \caption{Temporal evolution of the MEQ cloud’s spatial extent. The three rows correspond to the training (top), validation (middle), and test (bottom) datasets. In each row, the solid curve shows the observed 50th-percentile distance (\(P_{50}\)) and the dashed curve the observed 95th-percentile distance (\(P_{95}\)). Forecasts from the 1-, 15-, and 30-second models are plotted in blue, red, and green, respectively. Shaded regions denote \(\pm\sigma\) (standard deviation).}

    \label{fig:spatial_extent_history}
\end{figure}

Accurately forecasting the spatial evolution of MEQ clouds is critical for delineating the affected area, guiding mitigation, and optimizing future well placement \cite{boyet2024forecasting}. Figure \ref{fig:spatial_extent_history} compares the spatial extent of the MEQ clouds across the training, validation, and test sets, quantified by the 50th and 95th percentiles of the Euclidean distance from the injection point. The 1-second and 15-second forecast models reproduce the ground truth trajectories of both the median distance ($P_{50}$) and the far distance ($P_{95}$), achieving $R^2>0.97$ for the 1-second forecast model and $R^2>0.94$ for the 15-second forecast model. 

\begin{figure}[H]
    \centering
    \includegraphics[width=1.0\textwidth]{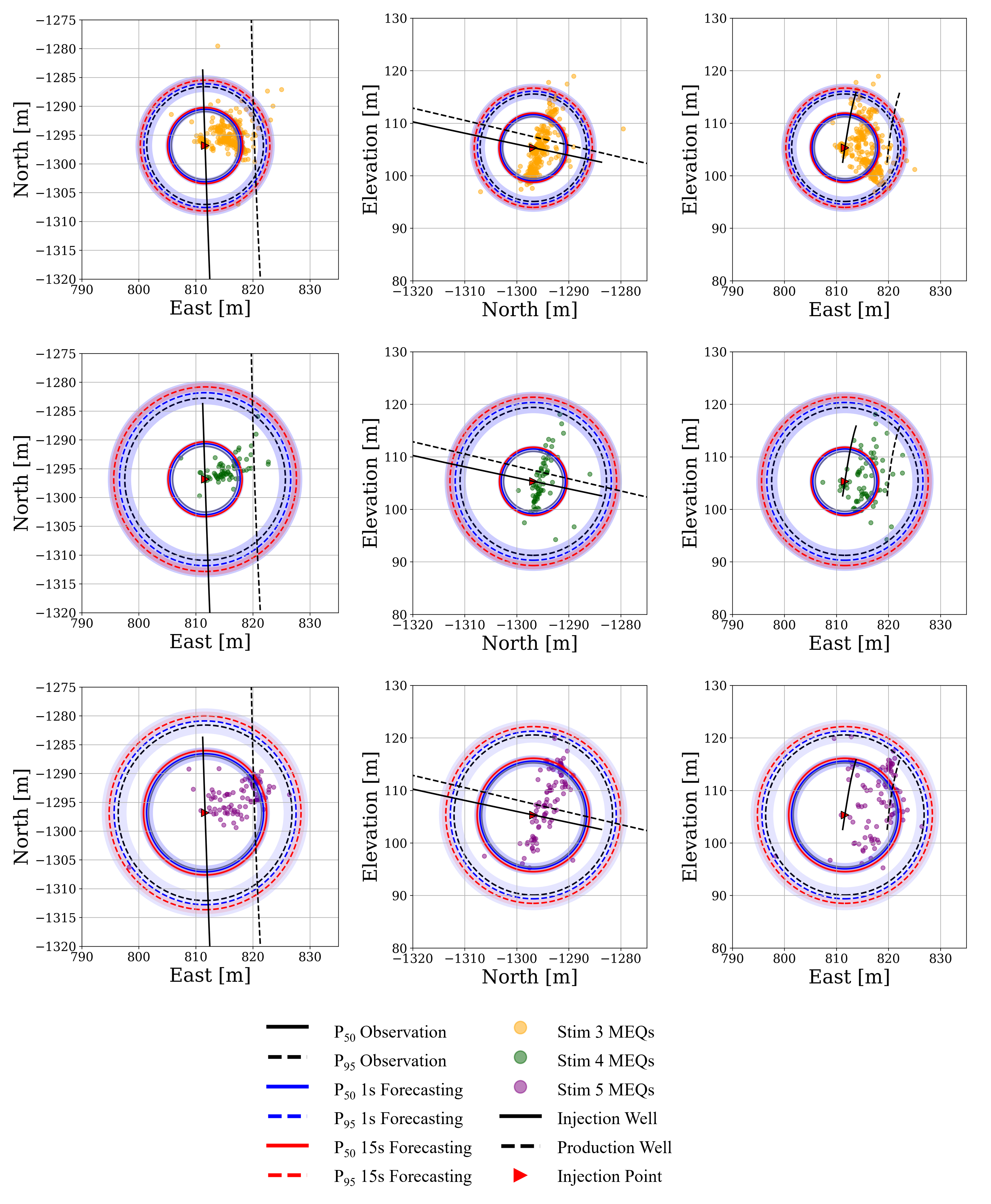}
    \caption{Spatial evolution of microearthquake (MEQ) clouds and forecast performance.  
    The figure shows the spatial distribution of MEQs and forecast results for different datasets and time horizons. Each row represents a dataset: training (top), validation (middle), and test (bottom). Each column shows projections on the XY, YZ, and ZX planes. Solid circles indicate the observed 50th-percentile radius ($P_{50}$), while dashed circles represent the 95th-percentile radius ($P_{95}$). Blue and red lines show forecasts from the 1-second and 15-second models, respectively, with shaded regions denoting $\pm \sigma$ (one predicted standard deviation). Colored dots represent MEQs from different stimulation phases: yellow for Stim 3, green for Stim 4, and purple for Stim 5. Solid black lines indicate the injection well, dashed black lines the production well, and red triangles mark the injection point. All spatial dimensions are in meters.}
    \label{fig:spatial_extent}
\end{figure}

Figure \ref{fig:spatial_extent} illustrates the final stabilized extents predicted by these models: absolute errors are below 0.4 m for the 1-second model and below 2 m for the 15-second model (Table \ref{tab:spatial_extent_summary}). For the 1-second case, the observed–predicted differences lie within the model’s $\pm \sigma$ band, indicating that the discrepancies are consistent with the reported uncertainty. In contrast, the 15-second differences exceed $\sigma$, revealing the limitations of the mid-range model. The 30-second model significantly underestimates both $P_{50}$ and $P_{95}$ in all data splits, highlighting its reduced reliability for long-range spatial forecasts.

\begin{table}[h]
    \centering
    \caption{\(R^{2}\) scores for all metrics and forecast models. The 1 s, 15 s, and 30 s columns correspond to models with forecast ranges of 1, 15, and 30 seconds, respectively.}
    \label{tab:R2_summary}
    \small
    \setlength{\tabcolsep}{6pt}
    \renewcommand{\arraystretch}{1.15}
    \begin{tabular}{lcccccccccccc}
        \toprule
        & \multicolumn{3}{c}{Cum. MEQ count}
        & \multicolumn{3}{c}{$\mathcal{M}$}
        & \multicolumn{3}{c}{$P_{50}$ distance}
        & \multicolumn{3}{c}{$P_{95}$ distance}\\
        \cmidrule(lr){2-4}\cmidrule(lr){5-7}\cmidrule(lr){8-10}\cmidrule(lr){11-13}
        Dataset & 1 s & 15 s & 30 s & 1 s & 15 s & 30 s & 1 s & 15 s & 30 s & 1 s & 15 s & 30 s\\
        \midrule
        Train & 0.994 & 0.963 & 0.807 & 0.996 & 0.933 & 0.748 & 0.969 & 0.974 & 0.444 & 0.967 & 0.974 & 0.556\\
        Val   & 0.984 & 0.929 & 0.649 & 0.993 & 0.878 & 0.546 & 0.972 & 0.941 & $-0.373$ & 0.989 & 0.947 & 0.208\\
        Test  & 0.993 & 0.972 & 0.809 & 0.993 & 0.935 & 0.765 & 0.995 & 0.988 & 0.401 & 0.973 & 0.973 & 0.543\\
        \bottomrule
    \end{tabular}
\end{table}

\begin{table}[h]
\centering
\caption{Final MEQ spatial extent. For each split and percentile, the observed radius is followed by the neural-network (NN) prediction, absolute difference \(|\text{Obs} - \text{Pred}|\), and model uncertainty \(\sigma\) for the 1-, 15-, and 30-second forecasting ranges.}
\label{tab:spatial_extent_summary}
\small
\setlength{\tabcolsep}{4pt}
\renewcommand{\arraystretch}{1.15}
\begin{tabular}{lcc|ccc|ccc|ccc}
\toprule
& & & \multicolumn{3}{c|}{1 s model} & \multicolumn{3}{c|}{15 s model} & \multicolumn{3}{c}{30 s model}\\
\cmidrule(lr){4-6}\cmidrule(lr){7-9}\cmidrule(lr){10-12}
Dataset & Pct & Obs [m] &
Pred [m] & Diff [m] & $\sigma$ [m]&
Pred [m] & Diff [m] & $\sigma$ [m]&
Pred [m] & Diff [m] & $\sigma$ [m] \\
\midrule
Train & $P_{50}$ & 5.92 & 6.29 & 0.38 & 0.67 & 6.59 & 0.68 & 0.20 & 2.56 & 3.35 & 0.79 \\
      & $P_{95}$ & 10.23 & 10.74 & 0.51 & 1.46 & 11.36 & 1.13 & 0.51 & 5.32 & 4.91 & 1.43 \\
\midrule
Val   & $P_{50}$ & 5.71 & 6.16 & 0.46 & 0.65 & 6.50 & 0.79 & 0.20 & 2.45 & 3.25 & 0.77 \\
      & $P_{95}$ & 14.08 & 15.01 & 0.93 & 2.04 & 16.02 & 1.94 & 0.71 & 7.61 & 6.47 & 1.99 \\
\midrule
Test  & $P_{50}$ & 9.92 & 10.25 & 0.34 & 1.11 & 10.79 & 0.87 & 0.34 & 4.29 & 5.62 & 1.32 \\
      & $P_{95}$ & 15.23 & 15.95 & 0.72 & 2.18 & 16.82 & 1.58 & 0.76 & 7.60 & 7.63 & 2.16 \\
\bottomrule
\end{tabular}
\end{table}

\section{Discussion and conclusion}\label{sec:discussion_conclusion}
Our transformer model accurately forecasts fluid-induced MEQs, capturing both their temporal evolution and spatial growth (Table \ref{tab:R2_summary}). This dual capability is novel; earlier studies focused mainly on temporal predictions \cite{yu2024crustal, li2022induced}. Reliable spatiotemporal forecasts are essential for estimating permeability changes and mitigating the risks associated with induced seismicity. In the following, we discuss how permeability enhancement can be inferred from monitoring data and model outputs, how fracture characteristics can be estimated, and the potentials and limitations of deep-learning-based forecasting for field-scale, fluid-induced earthquakes.

\subsection{Estimation of permeability enhancement}
Estimating permeability enhancement is a critical task in EGS, yet direct measurements are challenging in the subsurface. This limitation also applies to our study—we aim to understand how permeability evolves during hydraulic stimulation, but no direct measurements were available from the field experiment. Although the correlation between MEQs and permeability remains elusive \cite{kneafsey2025egs}, we derive a physically grounded rationale to indirectly estimate permeability using model outputs. Specifically, we apply the cubic law for permeability, which relates changes in fracture aperture to permeability change \cite{witherspoon1980validity, ouyang1993evaluation}:

\begin{equation} 
    \Delta k = \frac{\left(b_0 + \Delta b\right)^3}{12s} - \frac{b_0^3}{12s}
\end{equation}
where $\Delta k$ is the permeability change, $b_0$ is the initial fracture aperture, $\Delta b$ is the aperture change, and $s$ is the spacing between parallel fractures. Assuming that the initial aperture $b_0$ is negligible compared to the aperture change (i.e., $b_0 \ll \Delta b$), we approximate the permeability evolution as $\Delta k \approx \frac{\Delta b^3}{12 s}$. Given that the EGS Collab Experiment 1 aimed to establish fracture networks via hydraulic fracturing (i.e., tensile fractures) \cite{morris2018experimental, kneafsey2025egs}, we assume the seismic moment is linked to normal displacement by tensile opening. The equivalent moment $M_0$ for a tensile opening can be expressed as \cite{yu2024crustal}:

\begin{equation} 
    M_0 = 2GA\Delta u_n
\end{equation}
where $G$ is the shear modulus, $A$ is the area of the fracture, and $\Delta u_n$ is the normal displacement across the fracture. Assuming the area $A$ of the fracture is proportional to the aperture ($A \propto \Delta b$) \cite{olson2003sublinear}, we establish a direct proportionality between seismic moment $M_0$ and permeability change as \cite{ishibashi2016linking}:

\begin{equation} 
    \log{M_0} \propto \frac{2}{3}\log{\Delta k}
\end{equation}

With these scaling relationships, we infer that the overall logarithmic permeability increment is linearly proportional to the logarithmic seismic moment, though this assumption primarily holds during early stimulation, where the initial aperture is significantly smaller than the aperture increment (i.e., $b_0 \ll \Delta b$). 

During the first stimulation, the observed cumulative logarithmic seismic moment reaches $\approx 3$ (Figure \ref{fig:seismic_moment} left), implying a permeability increase of roughly two orders of magnitude. The 1-second forecast reproduces this estimate, whereas the 15-second forecast model overpredicts the moment by about one order of magnitude, and the 30-second forecast model underpredicts it by a similar amount. Because the cumulative seismic moment predicted by our network can be mapped directly to permeability changes, the model provides a practical, indirect means of tracking permeability evolution during hydraulic stimulation---though this mapping is valid only for the initial seismic‐moment range where the derivation’s assumptions hold.

\subsection{Inference of the fracture characteristics}
In fluid injection operations, we need to control the spatial extent of fracturing. As an example, in EGS fields, it is crucial to prevent MEQs from extending beyond the region between injection and production wells while enhancing permeability within this region through fracturing. Our model provides estimates of two spatial extents of MEQs: the 95th percentile distance ($P_{95}$) and the 50th percentile distance ($P_{50}$). $P_{95}$ represents the far extent of MEQs, while $P_{50}$ indicates the most active MEQ regions, which likely correspond to areas of greatest permeability increase due to fracture generation and re-opening.

The importance of tracking $P_{95}$ and $P_{50}$ becomes clear when the spatial extents from each stimulation are compared (Table \ref{tab:spatial_extent_summary}). From stimulation 3 (training) to stimulation 4 (validation), the observed $P_{95}$ grows by 3.85 m (from 10.23 to 14.08 m), while $P_{50}$ retreats by 0.21 m (from 5.92 to 5.71 m), indicating a slight shrinkage of the seismically active zone. Our 1-second forecast model reproduces these shifts almost exactly, predicting a 4.27 m increase in $P_{95}$ (from 10.74 to 15.01 m) and 0.13 m retreat in $P_{50}$ (from 6.29 to 6.16 m); all absolute errors fall within the 1-second forecast model's $\pm \sigma$ band. Between stimulation 4 (validation) and stimulation 5 (test), the observed $P_{95}$ increases by 1.15 m (from 14.08 to 15.23 m), whereas $P_{50}$ advanced by 4.21 m (from 5.71 to 9.92 m). The 1-second forecast model again captures these trends, predicting a 0.94 m rise in $P_{95}$ (from 15.01 to 15.95 m) and 4.09 m increase in $P_{50}$ (from 6.16 to 10.25 m). By accurately forecasting $P_{50}$ and $P_{95}$ in real time, the network enables practitioners to infer fracture propagation and activation, making it a practical tool for managing stimulation where direct measurements are not feasible.

\subsection{Potential and challenges of deep learning forecasting}
Among the various deep learning approaches, we chose the transformer model as our core architecture. The success of the transformer model is driven by several key factors. First, the self-attention mechanism allows the model to capture long-term dependencies \cite{vaswani2017attention, chen2023long, nie2022time}, which are crucial in fluid-induced seismicity, where MEQs are influenced by cumulative fluid injection, pore pressure changes, and perturbed in-situ stress conditions \cite{williams2008assessment}. In particular, fluid-induced seismicity often exhibits long time intervals between injection and seismicity. For instance, the largest earthquake (local magnitude 3.9) at the deep geothermal site GEOVEN in Vendenheim occurred more than six months after shut-in \cite{lengline2023largest}. The self-attention mechanism enables the model to weigh the importance of different input features over time, making it highly suited for sequential data \cite{zhou2021informer}.

Second, transformers excel at processing spatiotemporal data \cite{giuliari2021transformer}, which is vital for accurately predicting the spatial distribution of MEQs. This ability provides critical insights into fracture propagation \cite{gischig2015rupture} and fluid migration \cite{bhattacharya2019fluid}, both of which are key factors in assessing the effectiveness of hydraulic stimulation. The model’s performance in predicting the spatial extent of seismic events reflects its capacity to capture both the temporal and spatial dynamics of fluid injection-induced microseismicity. Third, the transformer's non-recurrent architecture allows it to handle irregular time series data \cite{chen2024contiformer}, a common occurrence in microseismic monitoring due to variable injection schedules and operational pauses. This flexibility enhances the model’s robustness across different stimulation phases and geological settings, making it adaptable to varying conditions and data availability—a common challenge in real-world geophysical applications.

While the model shows promising results, extending it to large-scale field operations introduces additional uncertainties due to unknown geological heterogeneity and the extended temporal dependencies inherent to fluid-induced seismicity. The data used in this study were collected from an intermediate-scale (10–20 m) experiment with comprehensive monitoring tools from the EGS Collab project \cite{kneafsey2018overview, kneafsey2019egs}. Such dense instrumentation may not be feasible in reservoir-scale engineering applications, raising questions about the model’s generalizability to less controlled, large-scale environments. One promising strategy for adapting deep learning forecasting techniques to larger-scale fluid-induced seismicity applications involves transfer learning with fine-tuning. Yu et al. \cite{yu2024crustal} recently demonstrated successful transferability between datasets from Utah FORGE and EGS Collab using appropriate fine-tuning methods. Although further fine-tuning will likely be required to adjust the model to larger operational scales, the fundamental assumption remains that the neural network model learns generalizable signal patterns associated with fluid-induced MEQs. Additionally, integrating uncertainty quantification into predictions becomes increasingly important given the higher uncertainty inherent in real-field-scale operations. By incorporating these strategies, along with judicious monitoring, transformer networks could be systematically validated and effectively implemented at larger scales. Future work could involve training and validating the model’s performance with field-scale fluid-induced seismic data and hydraulic stimulation histories, thus ensuring robustness in more complex geological settings.


In summary, despite limitations related to monitoring systems and scale, this study presents a novel approach for forecasting MEQs in response to fluid injection. The transformer model’s ability to predict both temporal and spatial evolution highlights its potential as a valuable tool in subsurface operations, offering significant improvements in safety and efficiency.

\section{Method: Transformer neural network architecture}\label{sec:method}
We employ a transformer neural network to forecast the spatiotemporal evolution of fluid-induced microearthquakes (MEQs). The attention mechanism captures dependencies in the monitoring time series, allowing the model to learn patterns across multiple temporal scales.
Figure \ref{fig:architecture} illustrates the overall architecture. Given a sequence of past monitoring data, the model predicts the future MEQ features. The following subsections describe data processing, network architecture, loss function, and hyperparameter tuning.

\subsection{Data preprocessing: crop and normalization}\label{sec:dataPreprocessing}
We first construct training segments by sliding a growing stimulation history across the cumulative time series and advancing the forecast horizon in non-overlapping blocks. The monitoring data at discrete time index $t$ are defined as:

\begin{equation}
    \bm{x}(t) = \left[x_{1}(t),\, x_{2}(t), \,..., \, \, x_{6}(t) \right]^T \in \mathbb{R}^M \, ,
\end{equation}
where the monitoring dimension $M = 6$ includes hydraulic stimulation features ---(1) flow rate ($x_1$) and (2) well head pressure ($x_2$)--- and spatiotemporal MEQ features ---(3) cumulative MEQ numbers ($x_3$), (4) $\log{M_0}$ ($x_4$), (5) 95th percentile distance ($x_5$), (6) 50th percentile distance)($x_6$).

The cropping procedure is controlled by two hyperparameters. The minimum history length $l_{\min}$ specifies the number of monitoring samples always available, and the forecast horizon $l_{\text{future}}$ specifies how many future steps are predicted at once. For a monitoring ending at $t_\text{end}$, the number of segments is 
\begin{equation}
    N = \frac{t_{end} - l_{\min}}{l_{\text{future}}}
\end{equation}

For each segment index $k \in \{0,\, ..., \,N-1\}$ the split time is set as
\begin{equation}
    t_k^{\text{split}} = l_{\min} + k l_{\text{future}}
\end{equation}
Thus, the cumulative monitoring input ($\bm{X}^{(k)}$) and the subsequent forecast window ($\bm{Y}^{(k)}$) are defined as:
\begin{align}
    \bm{X}^{(k)} &= \{\bm{x}(t) \, | \, 1 \leq t \leq t_k^{\text{split}} \} \in \mathbb{R}^{t_k^{\text{split}} \times M}\\
    \bm{Y}^{(k)} &= \{\bm{x}(t) \, | \,t_k^{\text{split}} < t \leq t_k^{\text{split}} + l_{\text{future}} \} \in \mathbb{R}^{t_k^{\text{future}} \times F}
\end{align}

where $F = 4$ corresponds to the forecasting MEQ features: (1) cumulative MEQ count, (2) $\log{M_{0}}$, (3) $P_{95}$, and (4) $P_{50}$. Each successive segment index $k$ advances the split by $l_{\text{future}}$, ensuring that the predicted time blocks $\bm{Y}^{(k)}$ are non-overlap and contiguous, while the input window grows monotonically. This approach yields continuous, leakage-free forecasting segments that can be applied in real time once at least $l_{\min}$ monitoring have been acquired (Figure~\ref{fig:forecastingSchematic}).

To fairly normalize the data without information leakage from future steps, normalization is applied individually to each input window $\bm{X}^{(k)}$. For each monitoring dimension $m \in \{1, ..., M\}$ and each segment $k$, we define the normalization using only the known input window as follows:

\begin{equation}
    \tilde{\bm{x}}_{m}^{(k)} = \frac{\bm{x}_{m}^{(k)} - \min\limits_{1\leq t \leq t_{k}^{\text{split}}} x_m(t)}{\max\limits_{1\leq t \leq t_{k}^{\text{split}}} x_m(t) - \min\limits_{1\leq t \leq t_{k}^{\text{split}}} x_m(t)}, \quad 1 \leq t \leq t_k^{\text{split}}
\end{equation}

The normalization parameters obtained from each input window $\bm{X}^{(k)}$are then consistently applied to scale the corresponding forecast window $\bm{Y}^{(k)}$. This ensures that normalization relies exclusively on information available at the prediction time, thus avoiding any data leakage from future observations.

\subsection{Neural network architecture}
Our transformer neural network architecture employs a multi-head attention mechanism designed to effectively capture temporal dependencies from variable length sequences. Given an input monitoring sequence $\bm{X}^{(k)}$, the multi-head attention layer processes the input as follows \cite{vaswani2017attention}:

\begin{equation} 
\text{Attention}(\mathbf{Q}, \mathbf{K}, \mathbf{V}) = \text{softmax}\left(\frac{\mathbf{Q}\mathbf{K}^\top}{\sqrt{d_k}}\right) \mathbf{V} \, ,
\end{equation}
where $\mathbf{Q} = \mathbf{X}^{(k)}\mathbf{W}_Q$, $\mathbf{K} = \mathbf{X}^{(k)}\mathbf{W}_K$, and $\mathbf{V} = \mathbf{X}^{(k)}\mathbf{W}_V$ are the query, key, and value matrices, respectively; $\mathbf{W}_Q$, $\mathbf{W}_K$, and $\mathbf{W}_V$ are learnable weight matrices; $d_k$ is the dimension of key vectors. 

Following the attention layer, a feed-forward network (FFN) \cite{svozil1997introduction} is applied independently to each time step. The FFN consists of two linear transformations with a Rectified Linear Unit (ReLU) activation function:

\begin{equation} 
    \text{FFN}(\mathbf{z}) = \text{ReLU}(\mathbf{z}\mathbf{W}_1 + \mathbf{b}_1)\mathbf{W}_2 + \mathbf{b}_2 \, ,
\end{equation}

where $\mathbf{z}$ denotes the input from the attention output, and $\bm{W}_1, \bm{W}_2, \bm{b}_1$, and $\bm{b}_2$ are learnable parameters. 

To enhance training stability, layer normalization and residual connections are applied after both attention and feed-forward layers. These ensure effective gradient propagation and prevent training instabilities.

After attention and feed-forward layers, global average pooling and dense layers reduce the sequence to a single vector, producing predictions for the forecasting window $\bm{Y}^{(k)}$. In particular, the model predicts both the mean ($\mu$) and log-variance ($\log \sigma^2$) of these forecasting MEQ features to quantify prediction uncertainty:

\begin{equation}
    \bm{y}_{\text{pred}} \in \mathbb{R}^{l_{\text{future}} \times 2F}, \quad \bm{y}_{\text{pred}}(t) = [\mu_{1} (t), \,..., \,\mu_{F}(t), \,\log \sigma^2_{1}(t), \, ..., \, \log \sigma^2_{F}(t)]
\end{equation}

The model is trained using the Adam optimizer \cite{kingma2014adam} with a heteroscedastic Gaussian negative 
log-likelihood (NLL) loss function \cite{nix1994estimating, lakshminarayanan2017simple}, augmented by a monotonicity penalty weighted by the hyperparameter ($\lambda$):

\begin{equation}
    \mathcal{L} = \text{NLL}(\mathbf{y}_{\text{true}}, \mathbf{y}_{\text{pred}}) + \lambda \,\text{Penalty}_{\text{mono}}
\end{equation}

The NLL explicitly measures the discrepancy between predictions and true values, accounting for predictive uncertainty. Given the predicted mean ($\mu_{\text{pred}}$ and log-variance ($\log \sigma^2_{\text{pred}}$), the NLL is defined as:

\begin{equation}
    \text{NLL}(\mathbf{y}_{\text{true}}, \mathbf{y}_{\text{pred}}) = \frac{1}{2NF} \sum_{i=1}^{N} \sum_{f=1}^{F} \left[ \frac{(y_{i,f}^{\text{true}} - \mu_{i,f}^{\text{pred}})^2}{\sigma_{i,f}^{2}} + \alpha \log \left(\sigma_{i,f}^{2}\right) \right],
\end{equation}

where $N$ is the number of time steps in the forecast window, $F$ is the number of MEQ target features, and $\alpha$ is the hyperparameter to discourage the model from inflating variance. This formulation captures both prediction accuracy and model confidence, penalizing over- or under-confident forecasts.

To enforce non-decrease for the cumulative term forecastings, a monotonicity penalty is applied to cumulative MEQ count and cumulative logarithmic seismic moment. The penalty is defined as:

\begin{equation}
    \text{Penalty}_{\text{mono}} = \sum_{t=2}^{T} \left| \min\left(0, \mathbf{y}^t_{\text{pred}} - \mathbf{y}^{t-1}_{\text{pred}} \right) \right| \,,
\end{equation}
where only the selected cumulative features are included in the penalty term.

Finally, all predictions are rescaled using the inverse of the normalization applied during preprocessing. The model performance is evaluated using the coefficient of determination ($R^2$):

\begin{equation} 
R^2 = 1 - \frac{\sum_{i=1}^{n} \left(Y_i - \hat{Y}_i\right)^2}{\sum_{i=1}^{n} \left(Y_i - \bar{Y}\right)^2} \end{equation}
where $Y$ includes the four spatiotemporal MEQ features.  

\subsection{Neural-network hyper-parameter tuning}
The transformer model is trained to forecast spatiotemporal MEQs from hydraulic-stimulation history and past MEQ responses.  
While network weights are learned automatically, several settings—loss-function coefficients, architectural widths, batch size, dropout rate, and penalty weights—must be chosen by the user \cite{feurer2019hyperparameter,yang2020hyperparameter}.  
Table~\ref{tab:fixed_hparams} lists the values that remain fixed in every experiment.

Two coefficients are tuned by grid search:  
\(\beta\) (the variance-regularisation weight inside the heteroscedastic Gaussian NLL term) and  
\(\lambda\) (the weight on the monotonic-increase penalty applied to cumulative MEQ count and cumulative seismic moment).  
For each forecast horizon \(l_{\text{future}}\in\{1,15,30\}\) models are trained with \(\beta,\lambda\in\{0.1,1.0,10.0\}\).  
Validation \(R^{2}\) scores identify the optimal pair \((\beta^{\star},\lambda^{\star})\); the corresponding results appear in Table~\ref{tab:best_hparams}.  

Short-horizon models---forecast windows of up to fifteen seconds--- achieve excellent accuracy; for example, the $l_{\text{future}} = 15$ model reaches $R^{2}_{\text{val}}=0.924$. As the horizon lengthens, performance degrades: at $n_{\text{future}}=30$ the best model attains $R^{2}_{\text{val}}=0.046$.  
The horizon-specific models reported in Table~\ref{tab:best_hparams} are used for all subsequent experiments.

\begin{table}[h]
    \centering
    \caption{Fixed architecture and training settings used in every experiment.}
    \label{tab:fixed_hparams}
    \renewcommand{\arraystretch}{1.1}
    \begin{tabular}{@{}ll@{}}
    \toprule
    Contents & Value \\ \midrule
    Monitoring features $M$ & 6 \\
    Forecasting features $F$ & 4 per timestep \\
    Number of multi-head attention & 4 heads\\
    Feed-forward width & 32 units (ReLU activation) \\
    Dropout rate & 0.30 \\
    Batch size & 16 \\
    Maximum epochs & 100 (with early stopping) \\ \bottomrule
    \end{tabular}
\end{table}

\begin{table}[h]
    \centering
    \caption{Optimal \((\beta,\lambda)\) and validation \(R^{2}\) for each prediction horizon.}
    \label{tab:best_hparams}
    \begin{tabular}{@{}ccccc@{}}
    \toprule
    $n_{\text{future}}$ & \(\beta^{\star}\) & \(\lambda^{\star}\) & \(R^{2}_{\text{val}}\) \\ \midrule
    1  & 0.1 & 0.1 & 0.990 \\
    15 & 10  & 0.1 & 0.924 \\
    30 & 10  & 0.1 & 0.046 \\ \bottomrule
    \end{tabular}
\end{table}

\section*{Acknowledgements}
JC gratefully acknowledges the support of the Ingenuity: Next Generation Nuclear Waste Disposal Internship program, funded by the U.S. Department of Energy, Office of Nuclear Energy, and Office of Spent Fuel and Waste Disposition. This work was supported by the US Department of Energy (DOE), the Office of Nuclear Energy, Spent Fuel and Waste Science and Technology Campaign, and by the US Department of Energy (DOE), under Contract Number DE-AC02-05CH11231 with Lawrence Berkeley National Laboratory. 

\section*{Data Availability Statement}
The code used in this study is available on GitHub at \url{https://github.com/jh-chung1/Transformer_MEQ_Forecasting}. The EGS Collab experiment's stimulation data and seismic catalog are available at \cite{GDR_Dataset_1229} and \cite{GDR_Dataset_1166}.

\section*{Authorship Statement}
\textbf{Jaehong Chung}: Conceptualization, Methodology, Investigation, Visualization, Writing---original draft, Review and editing.  
\textbf{Michael Manga}: Conceptualization, Investigation, Supervision, Review and editing.  
\textbf{Timothy Kneafsey}: Conceptualization, Investigation, Supervision, Review and editing.  
\textbf{Tapan Mukerji}: Investigation, Supervision, Review and editing.
\textbf{Mengsu Hu}: Conceptualization, Investigation, Funding acquisition, Project administration, Supervision, Review and editing.

\section*{Declarations}
\textbf{Conflict of Interest}  
The authors declare no conflict of interest.




\bibliographystyle{plain} 
\bibliography{references}  

\begin{thebibliography}{10}

\bibitem{anikiev2023machine}
Denis Anikiev, Claire Birnie, Umair bin Waheed, Tariq Alkhalifah, Chen Gu, Dirk~J Verschuur, and Leo Eisner.
\newblock Machine learning in microseismic monitoring.
\newblock {\em Earth-Science Reviews}, 239:104371, 2023.

\bibitem{aochi2021statistical}
Hideo Aochi, Julie Maury, and Thomas Le~Guenan.
\newblock How do statistical parameters of induced seismicity correlate with fluid injection? case of oklahoma.
\newblock {\em Seismological Society of America}, 92(4):2573--2590, 2021.

\bibitem{bachu2003sequestration}
Stefan Bachu and Jennifer~J Adams.
\newblock Sequestration of {CO}$_2$ in geological media in response to climate change: capacity of deep saline aquifers to sequester {CO}$_2$ in solution.
\newblock {\em Energy Conversion and management}, 44(20):3151--3175, 2003.

\bibitem{baisch2009investigation}
Stefan Baisch, Robert V{\"o}r{\"o}s, Ralph Weidler, and Doone Wyborn.
\newblock Investigation of fault mechanisms during geothermal reservoir stimulation experiments in the cooper basin, australia.
\newblock {\em Bulletin of the Seismological Society of America}, 99(1):148--158, 2009.

\bibitem{bergen2019machine}
Karianne~J Bergen, Paul~A Johnson, Maarten~V de~Hoop, and Gregory~C Beroza.
\newblock Machine learning for data-driven discovery in solid earth geoscience.
\newblock {\em Science}, 363(6433):eaau0323, 2019.

\bibitem{bhattacharya2019fluid}
Pathikrit Bhattacharya and Robert~C Viesca.
\newblock Fluid-induced aseismic fault slip outpaces pore-fluid migration.
\newblock {\em Science}, 364(6439):464--468, 2019.

\bibitem{boyet2024forecasting}
Auregan Boyet, V{\'\i}ctor Vilarrasa, Jonny Rutqvist, and Silvia De~Simone.
\newblock Forecasting fluid-injection induced seismicity to choose the best injection strategy for safety and efficiency.
\newblock {\em Philosophical Transactions A}, 382(2276):20230179, 2024.

\bibitem{camps2021deep}
Gustau Camps-Valls, Devis Tuia, Xiao~Xiang Zhu, and Markus Reichstein.
\newblock {\em Deep learning for the Earth Sciences: A comprehensive approach to remote sensing, climate science and geosciences}.
\newblock John Wiley \& Sons, Hoboken, NJ, 2021.

\bibitem{chen2024contiformer}
Yuqi Chen, Kan Ren, Yansen Wang, Yuchen Fang, Weiwei Sun, and Dongsheng Li.
\newblock Contiformer: Continuous-time transformer for irregular time series modeling.
\newblock {\em Advances in Neural Information Processing Systems}, 36, 2024.

\bibitem{chen2023long}
Zonglei Chen, Minbo Ma, Tianrui Li, Hongjun Wang, and Chongshou Li.
\newblock Long sequence time-series forecasting with deep learning: A survey.
\newblock {\em Information Fusion}, 97:101819, 2023.

\bibitem{chung2024prediction}
Jaehong Chung, Rasool Ahmad, WaiChing Sun, Wei Cai, and Tapan Mukerji.
\newblock Prediction of effective elastic moduli of rocks using graph neural networks.
\newblock {\em Computer Methods in Applied Mechanics and Engineering}, 421:116780, 2024.

\bibitem{damen2006health}
Kay Damen, Andr{\'e} Faaij, and Wim Turkenburg.
\newblock Health, safety and environmental risks of underground {CO}$_2$ storage--overview of mechanisms and current knowledge.
\newblock {\em Climatic Change}, 74(1):289--318, 2006.

\bibitem{ellsworth2019triggering}
William~L Ellsworth, Domenico Giardini, John Townend, Shemin Ge, and Toshihiko Shimamoto.
\newblock Triggering of the {P}ohang, {K}orea, earthquake ${M}_w$ 5.5 by enhanced geothermal system stimulation.
\newblock {\em Seismological Research Letters}, 90(5):1844--1858, 2019.

\bibitem{feng2024monitoring}
Zongcai Feng, Lianjie Huang, Benxin Chi, Kai Gao, Jiaxuan Li, Jonathan Ajo-Franklin, Douglas~A Blankenship, Timothy~J Kneafsey, EGS~Collab Team, et~al.
\newblock Monitoring spatiotemporal evolution of fractures during hydraulic stimulations at the first {EGS} collab testbed using anisotropic elastic-waveform inversion.
\newblock {\em Geothermics}, 122:103076, 2024.

\bibitem{feurer2019hyperparameter}
Matthias Feurer and Frank Hutter.
\newblock Hyperparameter optimization.
\newblock {\em Automated machine learning: Methods, systems, challenges}, pages 3--33, 2019.

\bibitem{fiori2023monitoring}
R{\'e}mi Fiori, J{\'e}r{\^o}me Vergne, Jean Schmittbuhl, and Dimitri Zigone.
\newblock Monitoring induced microseismicity in an urban context using very small seismic arrays: {T}he case study of the {V}endenheim {EGS} project.
\newblock {\em Geophysics}, 88(5):WB71--WB87, 2023.

\bibitem{fu2021close}
Pengcheng Fu, Martin Schoenball, Jonathan~B Ajo-Franklin, Chengping Chai, Monica Maceira, Joseph~P Morris, Hui Wu, Hunter Knox, Paul~C Schwering, Mark~D White, et~al.
\newblock Close observation of hydraulic fracturing at {EGS} {C}ollab {E}xperiment 1: {F}racture trajectory, microseismic interpretations, and the role of natural fractures.
\newblock {\em Journal of Geophysical Research: Solid Earth}, 126(7):e2020JB020840, 2021.

\bibitem{gischig2015rupture}
Valentin~S Gischig.
\newblock Rupture propagation behavior and the largest possible earthquake induced by fluid injection into deep reservoirs.
\newblock {\em Geophysical Research Letters}, 42(18):7420--7428, 2015.

\bibitem{giuliari2021transformer}
Francesco Giuliari, Irtiza Hasan, Marco Cristani, and Fabio Galasso.
\newblock Transformer networks for trajectory forecasting.
\newblock In {\em 2020 25th international conference on pattern recognition (ICPR)}, pages 10335--10342. IEEE, 2021.

\bibitem{hainzl2005detecting}
Sebastian Hainzl and Yosihiko Ogata.
\newblock Detecting fluid signals in seismicity data through statistical earthquake modeling.
\newblock {\em Journal of Geophysical Research: Solid Earth}, 110(B5), 2005.

\bibitem{hincks2018oklahoma}
Thea Hincks, Willy Aspinall, Roger Cooke, and Thomas Gernon.
\newblock Oklahoma's induced seismicity strongly linked to wastewater injection depth.
\newblock {\em Science}, 359(6381):1251--1255, 2018.

\bibitem{hummel2013nonlinear}
Nicolas Hummel and Serge~A Shapiro.
\newblock Nonlinear diffusion-based interpretation of induced microseismicity: A barnett shale hydraulic fracturing case study.
\newblock {\em Geophysics}, 78(5):B211--B226, 2013.

\bibitem{ishibashi2016linking}
Takuya Ishibashi, Noriaki Watanabe, Hiroshi Asanuma, and Noriyoshi Tsuchiya.
\newblock Linking microearthquakes to fracture permeability change: The role of surface roughness.
\newblock {\em Geophysical Research Letters}, 43(14):7486--7493, 2016.

\bibitem{jinqiang2021review}
Wang Jinqiang, Prabhat Basnet, and Shakil Mahtab.
\newblock Review of machine learning and deep learning application in mine microseismic event classification.
\newblock {\em Mining of Mineral Deposits}, 2021.

\bibitem{johann2018surge}
Lisa Johann, Serge~A Shapiro, and Carsten Dinske.
\newblock The surge of earthquakes in {C}entral {O}klahoma has features of reservoir-induced seismicity.
\newblock {\em Scientific reports}, 8(1):11505, 2018.

\bibitem{kingma2014adam}
Diederik~P Kingma.
\newblock Adam: A method for stochastic optimization.
\newblock {\em arXiv preprint arXiv:1412.6980}, 2014.

\bibitem{kneafsey2025egs}
Tim Kneafsey, Pat Dobson, Doug Blankenship, Paul Schwering, Mark White, Joseph~P Morris, Lianjie Huang, Tim Johnson, Jeff Burghardt, Earl Mattson, et~al.
\newblock The {EGS} {C}ollab project: Outcomes and lessons learned from hydraulic fracture stimulations in crystalline rock at 1.25 and 1.5 km depth.
\newblock {\em Geothermics}, 126:103178, 2025.

\bibitem{kneafsey2020egs}
Timothy~J Kneafsey, Doug Blankenship, Patrick~F Dobson, Joseph~P Morris, Mark~D White, Pengcheng Fu, Paul~C Schwering, Jonathan~B Ajo-Franklin, Lianjie Huang, Martin Schoenball, et~al.
\newblock The {EGS} collab project: Learnings from experiment 1.
\newblock In {\em Proceedings of the 45th Workshop on Geothermal Reservoir Engineering}, pages 10--12. Stanford University Stanford, CA, 2020.

\bibitem{kneafsey2019egs}
Timothy~J Kneafsey, Doug Blankenship, Hunter~A Knox, Timothy~C Johnson, Jonathan~B Ajo-Franklin, Paul~C Schwering, Patrick~F Dobson, Joseph~P Morris, Mark~D White, R~Podgorney, et~al.
\newblock {EGS} collab project: Status and progress.
\newblock In {\em Proceedings 44th Workshop on Geothermal Reservoir Engineering, Stanford University}, 2019.

\bibitem{kneafsey2018overview}
Timothy~J Kneafsey, Patrick Dobson, Doug Blankenship, Joe Morris, Hunter Knox, Paul Schwering, Mark White, Thomas Doe, W~Roggenthen, E~Mattson, et~al.
\newblock An overview of the egs collab project: field validation of coupled process modeling of fracturing and fluid flow at the sanford underground research facility, lead, sd.
\newblock In {\em 43rd Workshop on Geothermal Reservoir Engineering}, volume 2018, 2018.

\bibitem{GDR_Dataset_1229}
Hunter Knox, Pengcheng Fu, Paul Schwering, Christopher Strickland, Dorothy Linneman, Vince Vermeul, Jeff Burghardt, and Mathew Ingraham.
\newblock Egs collab experiment 1 stimulation data.
\newblock Geothermal Data Repository, Pacific Northwest National Laboratory, https://doi.org/10.15121/1651116, 2020.
\newblock Accessed: 2025-06-16.

\bibitem{kumazawa2013quantitative}
Takao Kumazawa and Yosihiko Ogata.
\newblock Quantitative description of induced seismic activity before and after the 2011 tohoku-oki earthquake by nonstationary etas models.
\newblock {\em Journal of Geophysical Research: Solid Earth}, 118(12):6165--6182, 2013.

\bibitem{kumazawa2014nonstationary}
Takao Kumazawa and Yosihiko Ogata.
\newblock Nonstationary etas models for nonstandard earthquakes.
\newblock 2014.

\bibitem{lakshminarayanan2017simple}
Balaji Lakshminarayanan, Alexander Pritzel, and Charles Blundell.
\newblock Simple and scalable predictive uncertainty estimation using deep ensembles.
\newblock {\em Advances in neural information processing systems}, 30, 2017.

\bibitem{lengline2023largest}
O~Lenglin{\'e}, J~Schmittbuhl, K~Drif, S~Lambotte, M~Grunberg, J~Kinscher, C~Sira, A~Schlupp, M~Schaming, H~Jund, et~al.
\newblock The largest induced earthquakes during the geoven deep geothermal project, strasbourg, 2018--2022: from source parameters to intensity maps.
\newblock {\em Geophysical Journal International}, 234(3):2445--2457, 2023.

\bibitem{li2023physics}
Ziyan Li, David~W Eaton, and J{\"o}rn Davidsen.
\newblock Physics-informed deep learning to forecast m\^{} max during hydraulic fracturing.
\newblock {\em Scientific Reports}, 13(1):13133, 2023.

\bibitem{li2022induced}
Ziyan Li, Derek Elsworth, Chaoyi Wang, and EGS-Collab.
\newblock Induced microearthquakes predict permeability creation in the brittle crust.
\newblock {\em Frontiers in Earth Science}, 10:1020294, 2022.

\bibitem{lu2021coupled}
Jianrong Lu and Ahmad Ghassemi.
\newblock Coupled {T}hermo--{H}ydro--{M}echanical--{S}eismic {M}odeling of {EGS} {C}ollab {E}xperiment 1.
\newblock {\em Energies}, 14(2):446, 2021.

\bibitem{manga2016increased}
Michael Manga, Chi-Yuen Wang, and Manoochehr Shirzaei.
\newblock Increased stream discharge after the 3 {S}eptember 2016 ${M}_w$ 5.8 {P}awnee, {O}klahoma earthquake.
\newblock {\em Geophysical Research Letters}, 43(22):11--588, 2016.

\bibitem{maniar2018machine}
Hiren Maniar, Srikanth Ryali, Mandar~S Kulkarni, and Aria Abubakar.
\newblock Machine-learning methods in geoscience.
\newblock In {\em SEG International Exposition and Annual Meeting}, pages SEG--2018. SEG, 2018.

\bibitem{mcclure2011investigation}
Mark~W McClure and Roland~N Horne.
\newblock Investigation of injection-induced seismicity using a coupled fluid flow and rate/state friction model.
\newblock {\em Geophysics}, 76(6):WC181--WC198, 2011.

\bibitem{metz2005ipcc}
Bert Metz, Ogunlade Davidson, HC~De~Coninck, Manuela Loos, and Leo Meyer.
\newblock {\em IPCC special report on carbon dioxide capture and storage}.
\newblock Cambridge: Cambridge University Press, Cambridge, 2005.

\bibitem{mital2024modeling}
Utkarsh Mital, Mengsu Hu, Yves Guglielmi, James Brown, and Jonny Rutqvist.
\newblock Modeling injection-induced fault slip using long short-term memory networks.
\newblock {\em Journal of Rock Mechanics and Geotechnical Engineering}, 2024.

\bibitem{morris2018experimental}
Joseph~P Morris, P~Fu, P~Dobson, Jonathan Ajo-Franklin, TJ~Kneafsey, H~Knox, D~Blankenship, MD~White, J~Burghardt, TW~Doe, et~al.
\newblock Experimental design for hydrofracturing and fluid flow at the {DOE} {EGS} collab testbed.
\newblock In {\em ARMA US Rock Mechanics/Geomechanics Symposium}, pages ARMA--2018. ARMA, 2018.

\bibitem{mousavi2022deep}
S~Mostafa Mousavi and Gregory~C Beroza.
\newblock Deep-learning seismology.
\newblock {\em Science}, 377(6607):eabm4470, 2022.

\bibitem{mousavi2016seismic}
S~Mostafa Mousavi, Stephen~P Horton, Charles~A Langston, and Borhan Samei.
\newblock Seismic features and automatic discrimination of deep and shallow induced-microearthquakes using neural network and logistic regression.
\newblock {\em Geophysical Journal International}, 207(1):29--46, 2016.

\bibitem{nie2022time}
Yuqi Nie, Nam~H Nguyen, Phanwadee Sinthong, and Jayant Kalagnanam.
\newblock A time series is worth 64 words: Long-term forecasting with transformers.
\newblock {\em arXiv preprint arXiv:2211.14730}, 2022.

\bibitem{nix1994estimating}
David~A Nix and Andreas~S Weigend.
\newblock Estimating the mean and variance of the target probability distribution.
\newblock In {\em Proceedings of 1994 ieee international conference on neural networks (ICNN'94)}, volume~1, pages 55--60. IEEE, 1994.

\bibitem{olson2003sublinear}
Jon~E Olson.
\newblock Sublinear scaling of fracture aperture versus length: an exception or the rule?
\newblock {\em Journal of Geophysical Research: Solid Earth}, 108(B9), 2003.

\bibitem{ouyang1993evaluation}
Zhihua Ouyang and Derek Elsworth.
\newblock Evaluation of groundwater flow into mined panels.
\newblock In {\em International Journal of Rock Mechanics and Mining Sciences \& Geomechanics Abstracts}, volume~30, pages 71--79. Elsevier, 1993.

\bibitem{petrillo2024fluids}
Giuseppe Petrillo, Takao Kumazawa, Ferdinando Napolitano, Paolo Capuano, and Jiancang Zhuang.
\newblock Fluids-triggered swarm sequence supported by a nonstationary epidemic-like description of seismicity.
\newblock {\em Seismological Research Letters}, 95(6):3207--3220, 2024.

\bibitem{qin2022forecasting}
Yan Qin, Ting Chen, Xiaofei Ma, and Xiaowei Chen.
\newblock Forecasting induced seismicity in oklahoma using machine learning methods.
\newblock {\em Scientific Reports}, 12(1):9319, 2022.

\bibitem{qin2024source}
Yan Qin, Jiaxuan Li, Lianjie Huang, Martin Schoenball, Jonathan Ajo-Franklin, Douglas Blankenship, Timothy~J Kneafsey, EGS~Collab Team, et~al.
\newblock Source mechanism of khz microseismic events recorded in multiple boreholes at the first egs collab testbed.
\newblock {\em Geothermics}, 120:102994, 2024.

\bibitem{rajesh2021characterization}
R~Rajesh and Harsh~K Gupta.
\newblock Characterization of injection-induced seismicity at north central {O}klahoma, {USA}.
\newblock {\em Journal of Seismology}, 25:327--337, 2021.

\bibitem{reichstein2019deep}
Markus Reichstein, Gustau Camps-Valls, Bjorn Stevens, Martin Jung, Joachim Denzler, Nuno Carvalhais, and F~Prabhat.
\newblock Deep learning and process understanding for data-driven earth system science.
\newblock {\em Nature}, 566(7743):195--204, 2019.

\bibitem{ritz2024pseudo}
Vanille~Ariane Ritz, Leila Mizrahi, Victor Clasen~Repoll{\'e}s, Antonio~Pio Rinaldi, Vala Hj{\"o}rleifsd{\'o}ttir, and Stefan Wiemer.
\newblock Pseudo-prospective forecasting of induced and natural seismicity in the hengill geothermal field.
\newblock {\em Journal of Geophysical Research: Solid Earth}, 129(3):e2023JB028402, 2024.

\bibitem{rothert2010passive}
Elmar Rothert and Stefan Baisch.
\newblock Passive seismic monitoring: mapping enhanced fracture permeability.
\newblock {\em Proc. World Geotherm. Congr}, pages 25--29, 2010.

\bibitem{rutqvist2012geomechanics}
Jonny Rutqvist.
\newblock The geomechanics of {CO}$_2$ storage in deep sedimentary formations.
\newblock {\em Geotechnical and Geological Engineering}, 30:525--551, 2012.

\bibitem{GDR_Dataset_1166}
Martin Schoenball, Jonathan Ajo-Franklin, Michelle Robertson, Todd Wood, Doug Blankenship, Paul Cook, Patrick Dobson, Yves Guglielmi, Pengcheng Fu, Timothy Kneafsey, Hunter Knox, Petr Petrov, Paul Schwering, Dennise Rempleton, Craig Ulrich, Jiaxuan Li, Lianjie Huang, Benxin Chi, Chet Hopp, and The EGS Collab~Team.
\newblock Egs collab experiment 1: Microseismic monitoring.
\newblock Geothermal Data Repository, Lawrence Berkeley National Laboratory, https://doi.org/10.15121/1557417, 2019.
\newblock Accessed: 2025-06-16.

\bibitem{schoenball2020creation}
Martin Schoenball, Jonathan~B Ajo-Franklin, Doug Blankenship, Chengping Chai, Aditya Chakravarty, Patrick Dobson, Chet Hopp, Timothy Kneafsey, Hunter~A Knox, Monica Maceira, et~al.
\newblock Creation of a mixed-mode fracture network at mesoscale through hydraulic fracturing and shear stimulation.
\newblock {\em Journal of Geophysical Research: Solid Earth}, 125(12):e2020JB019807, 2020.

\bibitem{shapiro2015fluid}
Serge~A Shapiro.
\newblock {\em Fluid-induced seismicity}.
\newblock Cambridge University Press, Cambridge, 2015.

\bibitem{svozil1997introduction}
Daniel Svozil, Vladimir Kvasnicka, and Jiri Pospichal.
\newblock Introduction to multi-layer feed-forward neural networks.
\newblock {\em Chemometrics and intelligent laboratory systems}, 39(1):43--62, 1997.

\bibitem{vaswani2017attention}
A~Vaswani.
\newblock Attention is all you need.
\newblock {\em Advances in Neural Information Processing Systems}, 2017.

\bibitem{wang2017induced}
Chi-Yuen Wang, Michael Manga, Manoochehr Shirzaei, Matthew Weingarten, and Lee-Ping Wang.
\newblock Induced seismicity in {O}klahoma affects shallow groundwater.
\newblock {\em Seismological Research Letters}, 88(4):956--962, 2017.

\bibitem{williams2008assessment}
Colin~F Williams, Marshall~J Reed, Robert~H Mariner, Jacob DeAngelo, and S~Peter Galanis.
\newblock Assessment of moderate-and high-temperature geothermal resources of the united states.
\newblock Technical report, Geological Survey (US), 2008.

\bibitem{witherspoon1980validity}
Paul~Adams Witherspoon, Joseph~SY Wang, K~Iwai, and John~E Gale.
\newblock Validity of cubic law for fluid flow in a deformable rock fracture.
\newblock {\em Water resources research}, 16(6):1016--1024, 1980.

\bibitem{yang2020hyperparameter}
Li~Yang and Abdallah Shami.
\newblock On hyperparameter optimization of machine learning algorithms: Theory and practice.
\newblock {\em Neurocomputing}, 415:295--316, 2020.

\bibitem{yeo2020causal}
IW~Yeo, Megan~RM Brown, S~Ge, and KK~Lee.
\newblock Causal mechanism of injection-induced earthquakes through the ${M}_w$ 5.5 pohang earthquake case study.
\newblock {\em Nature communications}, 11(1):2614, 2020.

\bibitem{yu2024crustal}
Pengliang Yu, Ankur Mali, Thejasvi Velaga, Alex Bi, Jiayi Yu, Chris Marone, Parisa Shokouhi, and Derek Elsworth.
\newblock Crustal permeability generated through microearthquakes is constrained by seismic moment.
\newblock {\em Nature communications}, 15(1):2057, 2024.

\bibitem{yu2021deep}
Siwei Yu and Jianwei Ma.
\newblock Deep learning for geophysics: Current and future trends.
\newblock {\em Reviews of Geophysics}, 59(3):e2021RG000742, 2021.

\bibitem{zeng2023transformers}
Ailing Zeng, Muxi Chen, Lei Zhang, and Qiang Xu.
\newblock Are transformers effective for time series forecasting?
\newblock In {\em Proceedings of the AAAI conference on artificial intelligence}, volume~37, pages 11121--11128, 2023.

\bibitem{zhai2019pore}
Guang Zhai, Manoochehr Shirzaei, Michael Manga, and Xiaowei Chen.
\newblock Pore-pressure diffusion, enhanced by poroelastic stresses, controls induced seismicity in {O}klahoma.
\newblock {\em Proceedings of the National Academy of Sciences}, 116(33):16228--16233, 2019.

\bibitem{zhang2022application}
Wengang Zhang, Xin Gu, Libin Tang, Yueping Yin, Dongsheng Liu, and Yanmei Zhang.
\newblock Application of machine learning, deep learning and optimization algorithms in geoengineering and geoscience: Comprehensive review and future challenge.
\newblock {\em Gondwana Research}, 109:1--17, 2022.

\bibitem{zhou2021informer}
Haoyi Zhou, Shanghang Zhang, Jieqi Peng, Shuai Zhang, Jianxin Li, Hui Xiong, and Wancai Zhang.
\newblock Informer: Beyond efficient transformer for long sequence time-series forecasting.
\newblock In {\em Proceedings of the AAAI conference on artificial intelligence}, volume~35, pages 11106--11115, 2021.

\bibitem{zhu2019phasenet}
Weiqiang Zhu and Gregory~C Beroza.
\newblock Phase{N}et: a deep-neural-network-based seismic arrival-time picking method.
\newblock {\em Geophysical Journal International}, 216(1):261--273, 2019.

\end{thebibliography}

\end{document}